\documentclass[journal]{IEEEtran}
\usepackage{graphicx}
\usepackage{epsfig}
\usepackage{epstopdf}
\usepackage{amsmath}
\usepackage{mathrsfs}
\usepackage{amssymb}
\usepackage{bm}
\usepackage{float}
\usepackage{indentfirst}
\usepackage{cite}
\usepackage{cases}
\usepackage{booktabs}
\usepackage{algorithmic}
\usepackage{algorithm}
\usepackage{multirow}
\usepackage{colortbl}
\usepackage{xcolor}
\usepackage[overload]{empheq}
\usepackage{MnSymbol}
\usepackage{underscore}
\usepackage[font = small]{caption}
\graphicspath{{./}} \DeclareGraphicsExtensions{.pdf,.jpg,.png}

\usepackage{array}
\makeatletter
\newcommand{\thickhline}{%
    \noalign {\ifnum 0=`}\fi \hrule height 1pt
    \futurelet \reserved@a \@xhline
}
\newcolumntype{"}{@{\hskip\tabcolsep\vrule width 1pt\hskip\tabcolsep}}
\makeatother
\definecolor{mygray}{gray}{0.7}

\definecolor{Gray}{gray}{0.85}
\definecolor{LightCyan}{rgb}{0.88,1,1}
\definecolor{mycyan}{cmyk}{.3,0,0,0}
\definecolor{mypink}{rgb}{.99,.91,.95}
\definecolor{ballblue}{rgb}{0.13, 0.67, 0.8}
\definecolor{brightgreen}{rgb}{0.4, 1.0, 0.0}
\definecolor{babyblue}{rgb}{0.54, 0.81, 0.94}
\definecolor{aureolin}{rgb}{0.99, 0.93, 0.0}
\newcolumntype{a}{>{\columncolor{Gray}}r}

\author{Zhiyuan Yang  and Chee-Wooi~Ten}

\begin{document}
\title{Cyber-Induced Risk Modeling for Microprocessor-Based Relays in Substations}
\maketitle
\begin{abstract}
Once critical substations are compromised,  attack agents can coordinate among their peers to plot for maximizing disruption using  local control devices. For defenders, it is critical to enumerate and identify all digital relays to determine the systemic risks. Any combination of disruptive switching via  the compromised relays can result in misoperation or immediate effect to the system. The resulting consequence of these attack's initial events would possibly incur cascading failure to a grid.
This paper quantifies the criticality of substation protective relays with the combination of the outage level and its corresponding severity risk index.  The proposed hypothesized outages are based on the type of protective relaying, bus configuration of a substation, and commonly implemented relaying schemes, such as bus differential, directional overcurrent and distance relays, are studied. This preliminary work also provides three approaches of determination in  probabilities for sensitivity analysis. The proposed risk indices are evaluated using IEEE test systems.
\end{abstract}

\section{Introduction}
\IEEEPARstart{T}{HE} implementation of remote access to substations for system management  has introduced residual risks with electronic intrusion possibilities  \cite{electronics}. In the recent cyber incidents in Ukraine, evidences have shown that the manipulation of electronic devices via remote-connected intrusion to the critical cyber assets has enabled attackers to plot for massive outages \cite{ukraine}. The attacks obviously were plausible because of the vulnerabilities allowing remote control capability through existing information and communications technologies that enable possible intrusion paths for attackers to access the network  \cite{CRS1}. Upon successful intrusion through a remote access point, the attackers can disrupt the communication channel, supervise the system status, monitor critical data, and covertly operate the local control system \cite{cybervul}.

To secure the substation automation  network, the latest security standards released by North American Electric Reliability Corporation (NERC)'s critical infrastructure protection (CIP) requires power company, utility, vendors, and other suppliers to regularly perform risk analysis on the critical assets in order to improve the system security \cite{CIP,IEEE234}. As the power communication infrastructure has now evolved to a more highly interconnected network, the National Institute of Standards and Technology (NIST) framework provides a guideline to systematically assess the risks, vulnerabilities, and securities of the smart grid technologies  \cite{NIST1}. Additionally, the IEEE Standard 1686 \cite{IEEE1686} is published to further emphasize the functions and features provided by the intelligent electronic devices (IEDs) should be incorporated in CIP program that can enhance the security level of these critical cyber assets.

The fundamental question remains -- what if a substation is compromised? What can happen to the entire power grid if attackers successfully execute their plan? There have been technical discussions about hypothetical impact scenarios by enumerating the combinations of potential catastrophe through unauthorized remote connection  \cite{5:Cyberbase}. This includes the preliminary investigation on the specific protective relaying scheme, bus differential configuration, and what would be the impact if the IED is compromised \cite{8:riskindex, IEDvul}. There still remains a challenge about how to effectively enumerate all hypothesized scenarios of cyber-induced attacks that can help defenders to identify pivotal combinations \cite{5:Cyberbase,8:riskindex,IEDvul, Enumeration1}. The widely implemented tool for steady-state approach \cite{17:matpower} has been an effective pre-screening methodology to identify problematic scenarios that can be initiated with electronic manipulation. The proof-of-concept of active command mediation defense (A*CMD) has been implemented to thwart intrusion \cite{subnetwork}.

\section{Risk Modeling of Relay Outages}
\begin{figure*}[!h]
  \centering
  \includegraphics[scale=0.52]{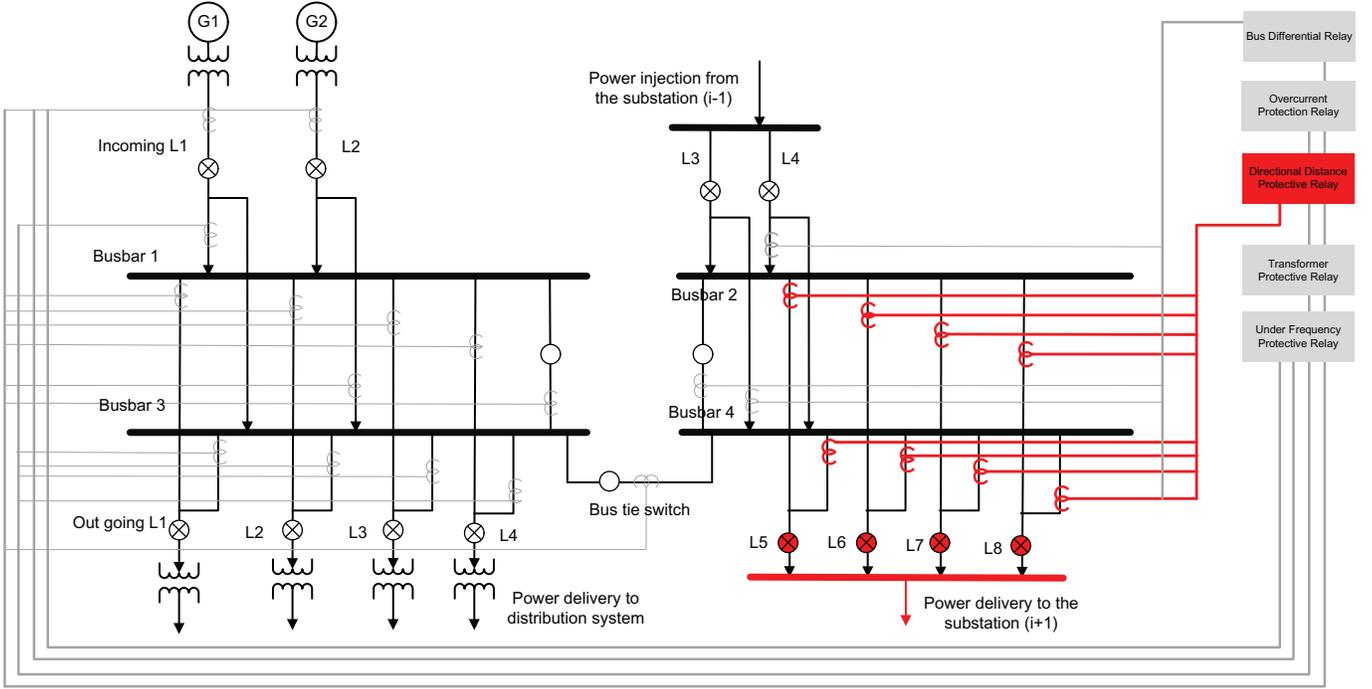}
  \caption{Schematic diagram of protective IEDs and the control perimeters within a substation}\label{Controllability}
\end{figure*}

The combinations of the substation criticality have been investigated in the recent years on  ``bottleneck list'' in \cite{5:Cyberbase} and \cite{8:riskindex}, which essentially assess the risk level for each substation outage by identifying the cases that are critically-weakened conditions based on presumed attack scenarios. With the consistent representation of terminology in the definition of ``critical'' cases, the section extends the modeling of relay outages that can be risky to manipulated that may initiate system-wide instability. Fig. \ref{Controllability} depicts the details of the connectivity relationship between the protective IEDs and their corresponding electrically controlling components in the substation $i$. Suppose that the substation $i$ has been compromised, the attackers would be able to manipulate single or multiple protective relays. Different relay would generate different outage when it has been compromised. As depicted in Fig. \ref{Controllability}, if the directional distance relay has been compromised, attackers can manipulate a disruptive switching command to electrically disconnect  substation $i+1$, thus other relays may misoperate causing the transmission line $(i,i+1)$ to be de-energized. Similarly, the implications may occur on other relays within a substation or other regional substations so as the resulting impacts to the power grid operation.

To quantify this problem, we introduce the variable $\mathcal{C}_{i,k}$ to denote the set of the electrical components, including all the lines, transformers, generator units, and loads, which are electronically controlled by the protective IED $k$, at the substation $i$. To enumerate all the possible successful cyber intrusions in the substation $i$, it generates $\sum_{k\in K} 2^{|\mathcal{C}_{i,k}|}$ different consequences, $K$ is the total number of the IEDs in the substation $i$. In our formulation, we'll assume the attacker would maximize the impacts by disconnecting as many electrical components as possible. Under this assumption, $\mathcal{C}_{i,k}'$ is the most ``severe'' set that is understudied in the proposed formulation, where $\mathcal{C}_{i,k}' \subset \mathcal{C}_{i,k}$.

Extended from the previous studies on the hypothetical substation outages, which reveal that the cyber-based contingency analysis is an enumerative study by extensively evaluating all the possible combinations of substations. The assessment of the digital protective relays is a much more complex problem, which includes a larger number of combinations:
\begin{equation}\label{Eq1}
\begin{split}
  \mathbb{S} & = (\sum_{k\in K_1} 2^{|\mathcal{C}_{1,k}|}) \cdot (\sum_{k\in K_2} 2^{|\mathcal{C}_{2,k}|}) \cdots (\sum_{k\in K_S} 2^{|\mathcal{C}_{S,k}|})  \\
  & = \prod_{i \in S} (\sum_{k\in K_i} 2^{|\mathcal{C}_{i,k}|})
  \end{split}
\end{equation}
where $S$ is the total number of the substations in the system and $K_i$ is the total number of the IEDs in the substation $i$. The constant $2$ indicates the open or closed status for each scenario of associated IEDs and substations, which implies $2^{|\mathcal{C}_{i,k}|}$ scenarios if IED $k$ is compromised. In our formulation, the most ``severe'' set $\mathcal{C}_{i,k}'$ is considered for each IED $k$, which would create $K_i$ different scenarios in total within single substation $i$. Thus, the Eq. \ref{Eq1} can be modified as:
\begin{equation}\label{Eq11}
    \mathbb{S} = 2^{K_1} \cdot 2^{K_2} \cdots 2^{K_S} = \prod_{i \in S} 2^{K_i} = 2^{\sum K_i}
\end{equation}
where $2^{K_i}$ represents the total number of outages of IEDs in substaion $i$. It is observed that $\mathbb{S}$ is determined by the configurations of IEDs in each substation or the total number IEDs in the system.

The right side of the Eq. \ref{Eq11} is the sum of the $S$-select-$k$ formulation, which is expressed as the designated outages of two or more components/substations and would produce different outcomes with or without considering outages of IEDs.  For instance, the $S$-select-$3$ problem on the substation outages would enumerate $C_{3}^{30} = 4,060$ cases in IEEE 30-bus system. However, from the Table \ref{T1}, it is observed that there are 106 IEDs are evaluated, which would generate $192,920$ scenarios consequently. It's predictable that the number of combinations would be greatly increased when studying larger cases. In this paper, we emphasize on detailing such hypothesized outages to the device level (the digital relays) to determine the  relay outages. Different relay types  may result in a consequentially different effect on the system than the attack on distance relays.

\begin{table}[!h]
\caption{Results of standard deviation of IEEE test systems}\label{T1}
\begin{tabular}{|l|l|l|l|c|}
  \hline
  \thickhline
 Test systems & $\sigma \le 1\%$ & $ 1\% < \sigma \leq 5\%$ & $ 5\% < \sigma \leq 10\%$  & Total \#. \\
\hline
 IEEE 30-bus & 1 & 49 & 34  & 106 \\

 IEEE 39-bus & 4 & 61 & 47 & 131 \\

 IEEE 57-bus & 7 & 74 &  71 & 245 \\

 IEEE 118-bus & 24 & 238 & 121 & 439 \\

 IEEE 300-bus & 43 & 411 & 354 & 959 \\
 \hline
\end{tabular}
\end{table}

\section{Cyber-Induced Impact Assessment }
\subsection{Probabilities and combinations}
A standard evaluation model for quantifying the risk of the disturbances or the outages is represented by the product of the event probability and its severity \cite{Risk-based}:

\begin{equation}\label{Eq2}
\mathbf{R}_{i,k} = \mathbf{Pr}_{i,k}\cdot\mathbf{Sr}_{i,k}
\end{equation}
where $\mathbf{R}_{i,k}$ is the risk index of the protective IED $k$ at substation $i$, $\mathbf{Pr}_{i,k}$ denotes the probability of event when the cyber intruder successfully hacks in the substation $i$ and manipulate the protective IED $k$, consequently, $\mathbf{Sr}_{i,k}$ is the severity of the outages, which, in this paper, is represented using the most ``severe'' set $\mathcal{C}_{i,k}'$ . To simulate probability of the intruding attempts, this paper assigns the probabilities based on the size of the set $\mathcal{C}_{i,k}'$. The Eq. \ref{Eq2} is elaborated as follows:

\begin{equation}\label{Eq3}
\mathbf{R}_{i,k} =
\begin{cases}
  \frac{|\mathcal{C}_{i,k}'|}{\sum_{k \in K} |\mathcal{C}_{i,k}'|}\cdot \frac{\sum_{i \in S}\sum_{k \in K} |\mathcal{P}_{i,k}|}{\sum_{k \in K} |\mathcal{P}_{i,k}|}, & \mbox{If  diverged}\\
  \frac{|\mathcal{C}_{i,k}'|}{\sum_{k \in K} |\mathcal{C}_{i,k}'|}\cdot \frac{|\mathcal{P}_{i,k}|}{\sum_{k \in K} |\mathcal{P}_{i,k}|}, & \mbox{Otherwise}
\end{cases}
\end{equation}
where  $\mathbf{Pr}_{i,k}$ = $|\mathcal{C}_{i,k}'|/\sum_{k \in K} |\mathcal{C}_{i,k}'|$ is the probability of the successful intruding attempts towards the protective relay $k$ in the substation $i$. In the study, it is assumed that an attacker does not acquire the knowledge of the power system and does not have a complete information of the entire power grid. They would enumerate all trials based on the connectivity degrees of the relays that connect more electrical components. These can be  represented by $\mathcal{C}_{i,k}'$. The probability is then calculated by measuring the proportion of the size of the ``severe'' set $|\mathcal{C}_{i,k}'|$ to the sum of the ``severe'' sets for all the protective relays.

In the Eq. \ref{Eq3}, the variable $\mathcal{P}_{i,k}$ denotes the total injected power to the substation node $i$, which are electronically controlled by the relay $k$. When the intruder successfully compromises the control panel and has the access to the IED $k$, all the power that is connected to this IED is considered as potential risks. To quantify the outage severity, the power flow evaluation is applied to verify the solutions of the study in which the outcome can be either converged or  diverged. For this reason, two different indices are proposed.

In the first scenario, if a solution agrees with a converged result, the severity of the outage $\mathbf{Sr}_{i,k}$ is determined by calculating the quotient by dividing the injected power that connected to the protective IED $k$ using the total injected power to the substation $i$, which is represented as $|\mathcal{P}_{i,k}| / \sum_{k \in K} |\mathcal{P}_{i,k}|$. The quotient locates the threshold [0,1]. If the solution is diverged, which suggests that the system can be unstable. Under this scenario, the severity of such event is considered to be much more severe where a potential system-wide blackout would occur. $\mathbf{Sr}_{i,k}$ = $\sum_{i \in S}\sum_{k \in K} |\mathcal{P}_{i,k}| / \sum_{k \in K} |\mathcal{P}_{i,k}|$ is quantified as the severity of the outage. Note that the numerator is the total power injection for the system. The proposed metric assures a comparable larger index than the previous conditions.

\subsection{Sensitivity analysis using standard deviation}
To evaluate the performance of the proposed metric, the assigned probability $\mathbf{Pr}_{i,k}$  captures successful intrusion of a given cyber network using pseudo-random numbers.  The results of the equal distribution of the probability are given in this problem, in order to initiate a comparison study.  The pseudo-random number is calculated through:
\begin{equation}\label{Eq4}
\mathbf{P}_{i,k} = \frac{\mathbf{P}_{i,k}'}{\sum_{k\in K} \textrm{rand}(1,K)}
\end{equation}
where rand$(1,K)$ represents the function that generates an array of random numbers within the interval (0,1), which has length of $K$. $\mathbf{P}_{i,k}'$ is the $k$-th elements in the array. The sum of these random number $\mathbf{P}_{i,k}'$ will not necessarily give a 1.0, which is not acceptable in probability distribution function. The scaling process is given by dividing each random number with the sum of all the random numbers, thus, the modified variable $\mathbf{P}_{i,k}$ is the $k$-th scaled random number which can be used as the probability assigned to the outage of the relay $k$ in the substation $i$.

\begin{figure}[!t]
  \centering
  \includegraphics[scale=0.46]{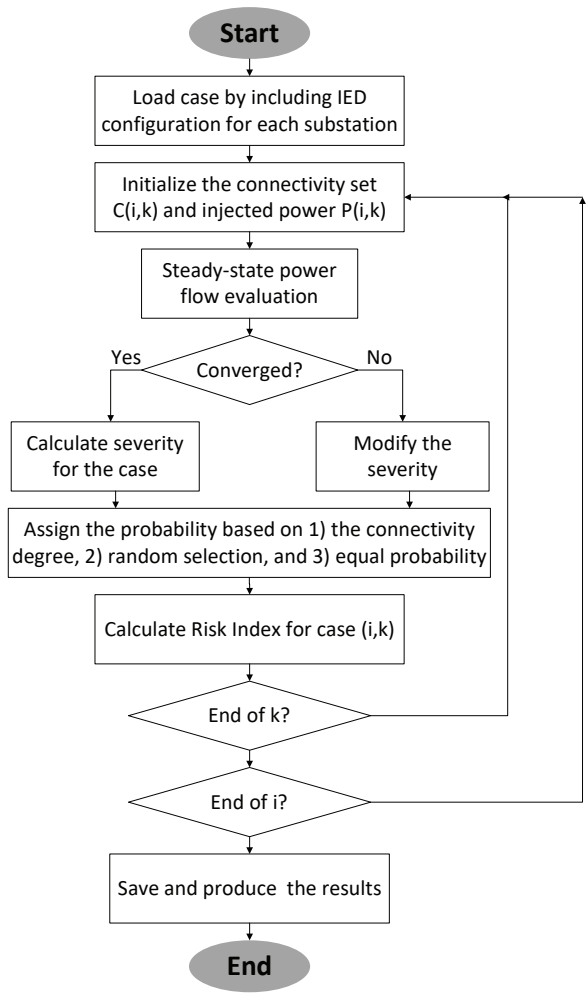}
  \caption{Algorithmic enumeration of relay outages}\label{Flowchart}
\end{figure}

Accordingly, the standard deviation $\sigma$ is given as the index to assess the performance of the proposed metric.
\begin{equation}\label{Eq5}
3\sigma_{i,k}^2 = (\mathbf{R}_{i,k}^{C}-\mathbf{R}_{i,k}^{A})^2+(\mathbf{R}_{i,k}^{R}-\mathbf{R}_{i,k}^{A})^2+(\mathbf{R}_{i,k}^{E}-\mathbf{R}_{i,k}^{A})^2
\end{equation}
where $\mathbf{R}_{i,k}^{C}$ denotes the risk of protective IED $k$ in the substation $i$ using the probability which are derived from the connectivity set $\mathcal{C}_{i,k}'$. Similarly, $\mathbf{R}_{i,k}^{R}$ and $\mathbf{R}_{i,k}^{E}$ are the risk indices using probabilities of pseudo random value and equal distribution method, correspondingly. $\mathbf{R}_{i,k}^{A}$ is the average risk index. $\sigma_{i,k}$ is the standard deviation of protective IED $k$ which is in substation $i$. Fig. \ref{Flowchart} describes the algorithm of the enumerative assessment for the protective relays. The proposed algorithm includes two loops, which include the iteration of power substations and the protective IEDs in each substation.

\section{Simulation Study}
\subsection{Results of hypothesized relay outages }
This study is validated using IEEE test systems. IEEE 30-bus system contains 6 generation units and 20 loads, which generate 191.6 MW power to the system and consume 189.2 MW power respectively. IEEE 39-bus system contains 9 generation nodes and 21 loads are connected to the system. 12 power transformers are connected to the system. It is observed that 6297.9 MW power is injected into the system and 6254.2 MW power is dispatched to the loads. IEEE 57-bus system is installed with 17 transformers and 7 generation units, which supply 1278.7 MW to the grid. The fixed 42 loads consume 1250.8 MW power in total. IEEE 118-bus system contains 54 generation units, 99 fixed loads, and 9 power transformers. The total power injection to the grid is 4374.9 MW and the total load consumption is 4242 MW power. 69 generators are installed in the IEEE 300-bus system with 23935.4 MW power supply and 201 loads totally consume 23525.8 MW of power. For each IEEE test case, it is initialized with five default protective relays in each substation, which are directional overcurrent relay, bus differential relay, directional distance relay, under frequency relay, and transformer relay. For different bus type and configuration, the set of protective relays will diversify. It is noticed that the in the Eq. \ref{Eq3}, the diverged solution would give a comparably higher value than the converged solution, in this section, in order to explicitly identify the ``worst'' cases, all the risk indices of diverged solutions are modified to 1.0.

\begin{figure}[!t]
  \centering
  \includegraphics[scale=0.26]{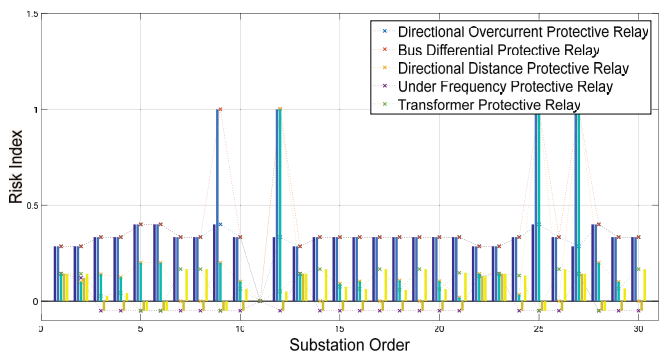}
  \caption{Risk index results of IEEE 30-bus system}\label{r1}
\end{figure}
\begin{figure}[!t]
  \centering
  \includegraphics[scale=0.26]{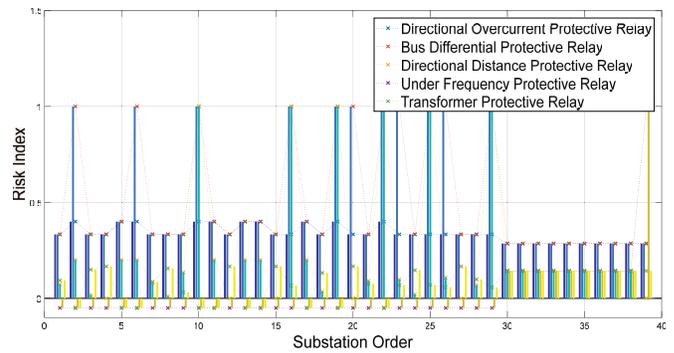}
  \caption{Risk index results of IEEE 39-bus system}\label{r2}
\end{figure}

\begin{figure}[!t]
  \centering
  \includegraphics[scale=0.26]{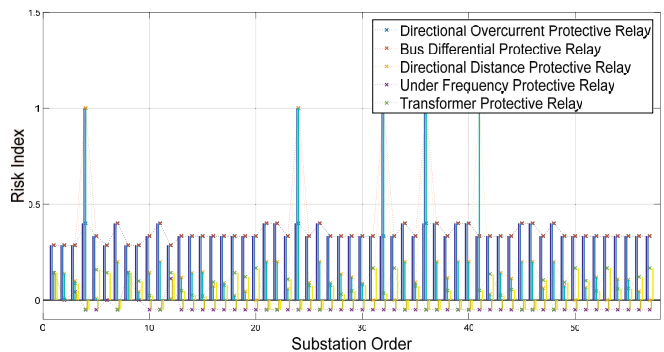}
  \caption{Risk index results of IEEE 57-bus system}\label{r3}
\end{figure}

\begin{figure}[!t]
  \centering
  \includegraphics[scale=0.26]{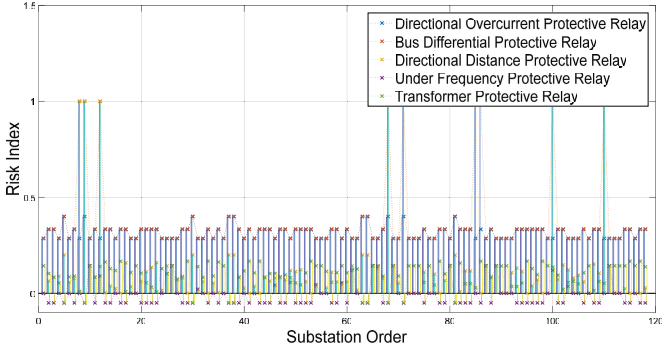}
  \caption{Risk index results of IEEE 118-bus system}\label{r4}
\end{figure}

\begin{figure}[!t]
  \centering
  \includegraphics[scale=0.122]{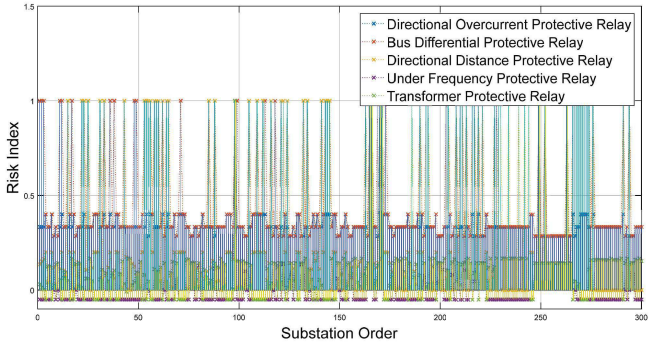}
  \caption{Risk index results of IEEE 300-bus system}\label{r9}
\end{figure}

Figs. \ref{r1}--\ref{r9} display the results of propsed risk index for different relays in IEEE 30-, 39-, 57-, 118-, and 300-bus systems, respectively. From these figures, the negative markers represent ``not available'' for such relay. For example, it is assumed in this study that the under frequency relays are equipped with the substations with generators. For those substations which are not classified as generation bus or load bus, the transformer relays are not equipped. In this respect, the negative risk value is given to differentiate the relay configurations between each substations. In the Fig. \ref{r1}, the bus 11 is not modeled with any relay as the solutions of steady-state analysis reveal that the power flows from/to the bus 11 is 0 MW, which, based on the definition in the Eq. \ref{Eq3}, would assign 0 as the risk index for each relay. However, the potential risks of relay outages in this substation may still exist. The cascading studies and transient analysis can be included to improve the existing model for future enhancement.

From the figures, it is observed that risk index for most relays are within [0.35, 0.50]. The critical IEDs are marked out with risk index of 1.0. For example, in IEEE 30-bus system, the overcurrent relay in the substation 9, bus differential and distance relays in the substation 12, 25, and 27, are identified as ``worst''  relays which would cause the system-wide instability in the steady-state analysis. In IEEE 39-bus system, it is observed that 18 out of 195 relays are identified as the critical relays, 11 out of which are bus differential relay. In IEEE 57-bus system, it can be observed that 9 out of 285 relays are evaluated as critical relays with 1.0 risk index, 5 of them are bus differential relays and 4 of them are distance relays. In IEEE 118-bus system, 16 out 590 relays are found to be critical and half of them are bus differential relays. Similarly, in the IEEE 300-bus system, 125 relays are identified as critical. 28\% of the relays are directional distance relays, 59\% of them are bus differential relays.


Generally, the distance relays in large-sized system are ranked as critical outage whereas the impacts of a relatively smaller system may not have similar impacts. For example, substation 186 is connected with a load demand of -21 MW, which would provide 21 MW to the grid. Once the distance relay has been compromised, the outgoing lines of the substation would be disconnected, in which case, all the branches (93-186, 185-186) adjacent to that bus would be consequently disconnected. Thus, the substation 186 is islanded and the unbalance between the generation and load demand cannot be absorbed in the steady-state power flow analysis. From the observation of this simulation study, the larger cases with a larger lumped load per location (substation) can also result in a higher risk level of distance relays.

By integrating the results found risk index evaluations, it is concluded that the bus differential relay acquires the highest risk index compared with other relays because 1) bus differential relay is the most commonly deployed relay in the substation and 2) it would electrically disconnect all the switches from the system if it has been compromised, which would change the system configuration and remove the substation from the initial setup in the test system. To improve the risk metric of the bus differential relays, the potential cascading failure is needed to be further studied. Additionally, the impacts of directional distance relay are higher  than the directional overcurrent relay due to the physical relations and relays where the disconnects affect abrupt change of the operating states. Compared with distance relay, overcurrent relay is assumed installed on the incoming lines from generators and local loads. When it has been compromised, the power injections and load demand would largely be disconnected from the system. However, the compromised distance relay would change the topology of the initial grid and consequently, cause substation islanded from the system, which is unstable in the study of steady-state evaluation.

\begin{figure}[!t]
  \centering
  \includegraphics[scale=0.26]{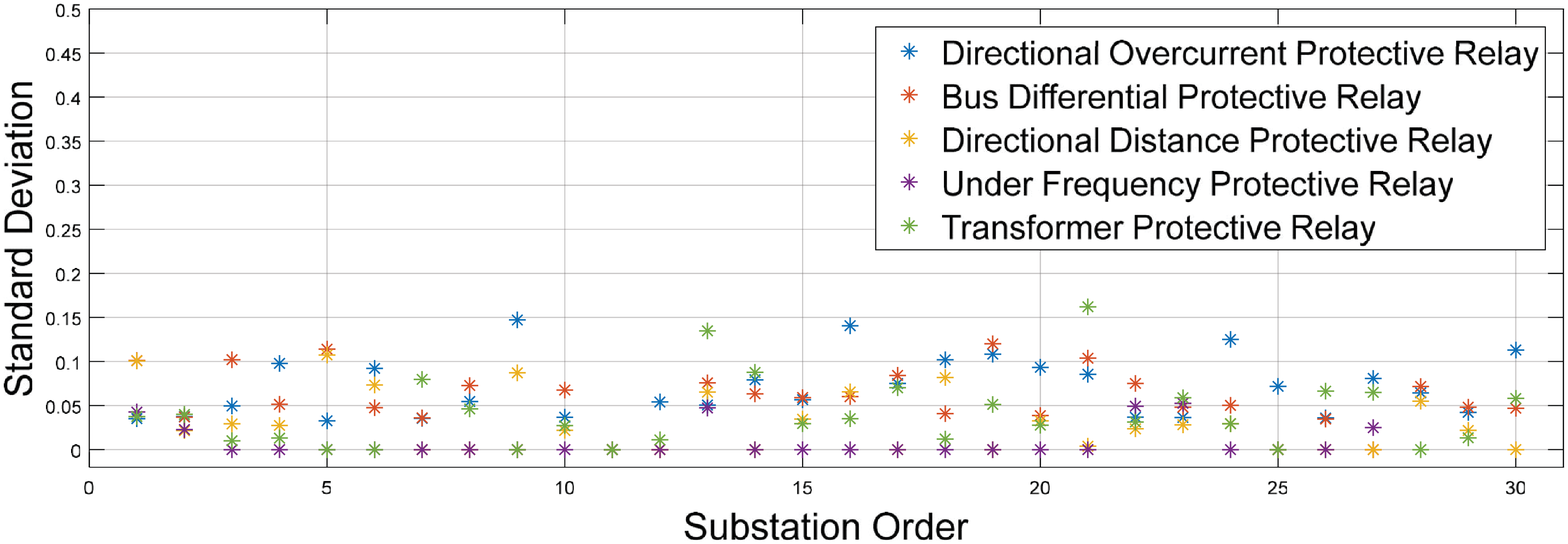}
  \caption{Standard deviation $\sigma$ for IEEE 30-bus test system}\label{r5}
\end{figure}

\begin{figure}[!t]
  \centering
  \includegraphics[scale=0.26]{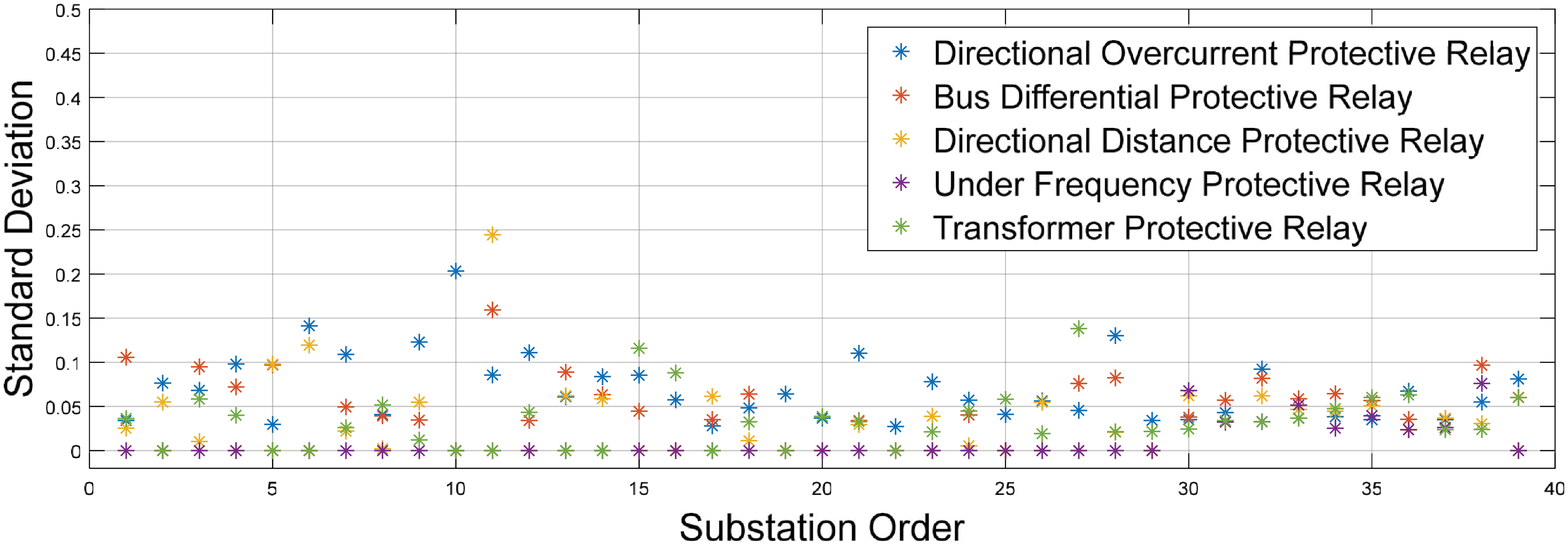}
  \caption{Standard deviation $\sigma$ for IEEE 39-bus test system}\label{r6}
\end{figure}

\subsection{Sensitivity analysis of proposed metric}
In order to compare the performance using different probability distribution, the Eq. \ref{Eq5} introduces the standard deviation to evaluate the sensitivity of the proposed metric. Table \ref{T1} in section II summarizes the detailed results of standard deviations for each IEEE test cases. The $\sigma$ is denoted by the standard deviation and the middle three columns record the number of relays whose $\sigma$ are in the thresholds of (0, 0.01], (0.01, 0.05], and (0.05, 0.1], respectively. Each substation might have different protective IED configuration. When starting the standard deviation evaluation, the relays that are recorded in negative risk index in the previous sections should be eliminated. For example, in the IEEE 30-bus system, under frequency relay is not available in the substation 3 but is equipped in the substation 2.

\begin{figure}[!t]
  \centering
  \includegraphics[scale=0.26]{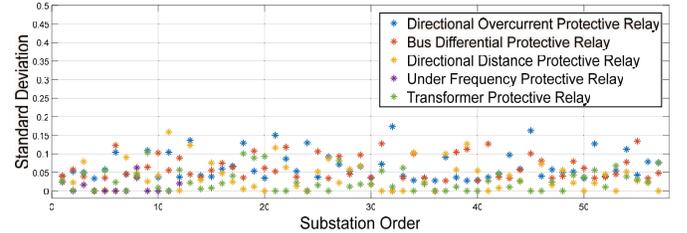}
  \caption{Standard deviation $\sigma$ for IEEE 57-bus test system}\label{r7}
\end{figure}

\begin{figure}[!t]
  \centering
  \includegraphics[scale=0.26]{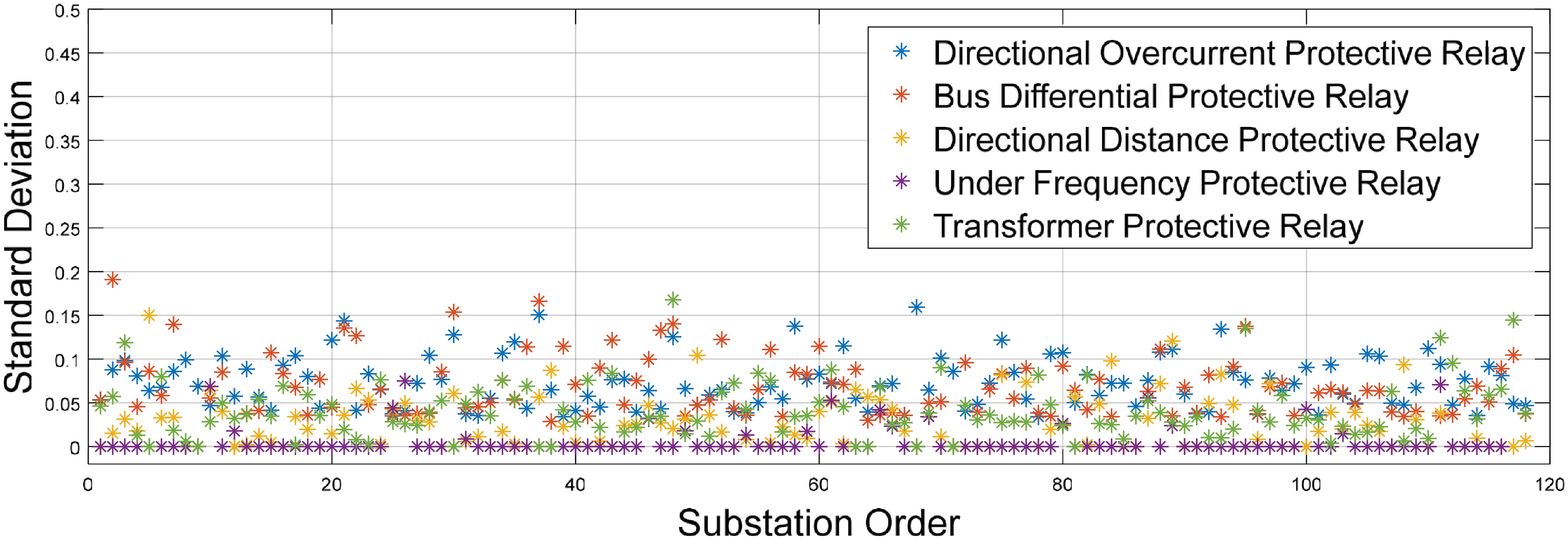}
  \caption{Standard deviation $\sigma$ for IEEE 118-bus test system}\label{r8}
\end{figure}

\begin{figure}[!t]
  \centering
  \includegraphics[scale=0.26]{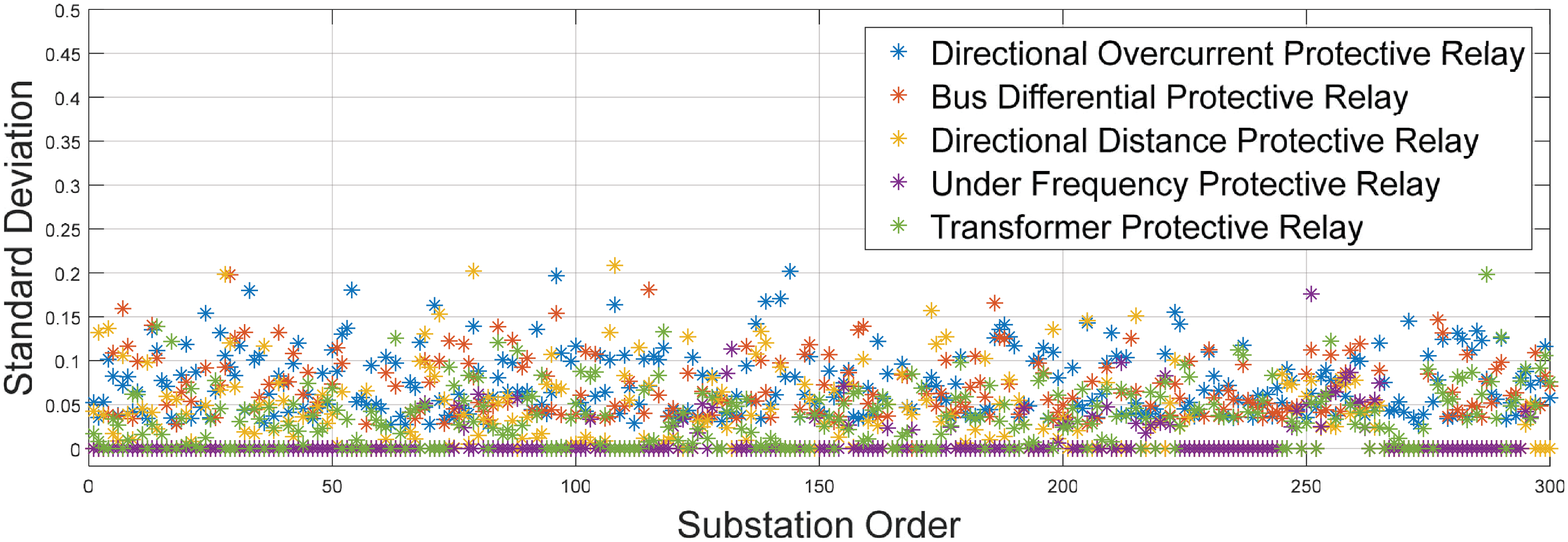}
  \caption{Standard deviation $\sigma$ for IEEE 300-bus test system}\label{r10}
\end{figure}

Figs. \ref{r5}--\ref{r10} show the standard deviations for different relays using different probability distributions in the IEEE test systems. Generally, it is recorded that 79\%, 85\%, 62\%, 87\%, and 84\% of relays acquire the standard deviation within 0.1 for IEEE 30-, 39-, 57-, 118-, 300-bus system, respectively.

\begin{figure}[!t]
  \centering
  \includegraphics[scale=0.26]{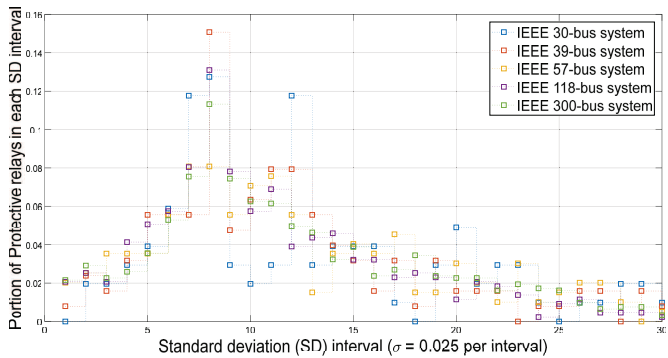}
  \caption{Statistical summary of the protective relays distribution according to standard deviation interval }\label{r11}
\end{figure}
As summarized in Table \ref{T1}, it is observed that the $\sigma$ of 80\% of relays are located in the interval (0, 0.1], which equals to the variance in the interval (0, 0.01]. Additionally, it is realized that the $\sigma$ for most scenarios are located in the section $3\% \le \sigma \leq 8\%$, which is the middle part of the in the interval (0, 0.1]. To specify the distribution,  Fig. \ref{r11} details the numerical results the relay distributions according to the standard deviation interval, notice that each big interval would represent 0.025 incremental of the standard deviation as the x-axis variable. The portion of the relays in the corresponding standard deviation interval is given as the y-axis variable. For example, combining these five test systems, approximate 25 \% of relays would locate in the first interval $\sigma \le 2.5\%$ and 30\% of relays are found in the interval $ 2.5\% \le \sigma \le 5\%$. According to the distribution sample points in the fig. \ref{r11}, it is statistically observed that such distribution can be fitted using a Normal distribution or Poisson distribution with the mean approximately equals to 8 units, which suggests that $\sigma$ equals to 0.04. The fitting function would be determined to calculate the confidence interval for the standard deviation. Additionally, it is revealed that the $\sigma$ of critical relays are much less than other relays. Because the Eq. \ref{Eq3} suggests that the higher the severity of the event, the lesser the risk index would be that is affected by the probability distribution. In this respect, the critical protective relays recorded in the risk index figures can be considered as reliable in the sense of steady-state analysis.

\section{Concluding Remarks}
The proposed framework further extends our worst-case assumptions of substation outage to the detailed level of abstraction at a device level, which is coupled to critical cyber assets within a substation. The analysis of impact level for hypothetical scenarios of substation protective relay outages is verified using the proposed metric. The metric quantifies the cyber-physical systemic risks of electronic intrusion. The combinatorial enumeration provides an extended exploration of potential attack strategies that is associated with different part of substation automation framework that is connected to electrical switchgear in physical facilities. To evaluate the performance of the risk-based analysis and improve the reliability of the index, we applied a comparison study by providing several probability distribution functions and investigate the standard deviation of the proposed model. The results of risk-based analysis have been validated using IEEE test systems by systematically enumerating hypothesized relay outages that may cause widespread instability. Future studies include improvements of multiple relay types in the enumeration and sequential permutation enumeration. This can also be incorporated as part of the substation outage enumeration and optimization that would effectively identify the critical scenarios that can help asset owners for planning purpose.

\bibliographystyle{IEEEtran}
\bibliography{IEEEabrv,RefDatabase}
\end{document}